\documentclass[12pt]{article}
\usepackage{amssymb}

\input{tcilatex}
\begin{document}

\title{{\LARGE Quantum motion in superposition of Aharonov-Bohm with some
additional electromagnetic fields}}
\author{V.G. Bagrov\thanks{%
Department of Physics, Tomsk State University, 634050, Tomsk, Russia. Tomsk
Institute of High Current Electronics, SB RAS, 634034 Tomsk, Russia; e-mail:
bagrov@phys.tsu.ru} , D.M. Gitman\thanks{%
Institute of Physics, University of Sao Paulo, Brazil; e-mail:
gitman@dfn.if.usp.br}, and A.D. Levin\thanks{%
Institute of Physics, University of Sao Paulo, Brazil; e-mail:
alevin@if.usp.br}}
\maketitle

\begin{abstract}
The structure of additional electromagnetic fields to the Aharonov-Bohm
field, for which the Schr\"{o}dinger, Klein-Gordon, and Dirac equations can
be solved exactly are described and the corresponding exact solutions are
found. It is demonstrated that aside from the known cases (a constant and
uniform magnetic field that is parallel to the Aharonov-Bohm solenoid, a
static spherically symmetrical electric field, and the field of a magnetic
monopole), there are broad classes of additional fields. Among these new
additional fields we have physically interesting electric fields acting
during a finite time, or localized in a restricted region of space. There
are additional time-dependent uniform and isotropic electric fields that
allow exact solutions of the Schrodinger equation. In the relativistic case
there are additional electric fields propagating along the Aharonov-Bohm
solenoid with arbitrary electric pulse shape.
\end{abstract}

\section*{Introduction}

The Aharonov-Bohm (AB) effect plays an important role in quantum theory
revealing a peculiar status of electromagnetic potentials in the theory.
This effect was discussed in \cite{AhaBo59} when studying the scattering of
a non-relativistic charged spinless particle by an infinitely long and
infinitely thin magnetic field of a solenoid (the AB field in what follows)
of finite magnetic flux (a similar effect was discussed earlier by Ehrenberg
and Siday \cite{EhrSi49}). The Schr\"{o}dinger equation with such a field
was exactly solved and it was found that a particle wave function vanishes
on the solenoid line. Although the particle does not penetrate the solenoid,
while the magnetic field vanishes outside of\emph{\ }it, the partial
scattering phases are proportional to the magnetic flux (modulo a flux
quantum) \cite{WuYa75}. A nontrivial particle scattering by such a field was
interpreted as a capability of a quantum particle to \textquotedblright
feel\textquotedblright\ an electromagnetic field vector potential because
the AB field vector potential does not vanish outside of the solenoid%
\footnote{%
It should be mentioned that in the relativistic case (Dirac equation with AB
field) some of wave functions from a complete set of solutions do not vanish
on the solenoid line.}. Solutions of the Dirac equations with AB field were
found and studied in detail, see e.g. \cite%
{PerVi89,Hagen91,Hagen93,VGS91,CNP92,CouPe94,FalPi01}.

A splitting of Landau levels in a superposition of the AB field and a
parallel uniform magnetic field (we call such a superposition the
magnetic-solenoid field-MSF) gives an example of the AB effect for bound
states. Solutions of the non-relativistic Schr\"{o}dinger equation with MSF
were first studied in \cite{Lewis83}. Solutions of the Klein-Gordon and
Dirac equations with MSF were first obtained in \cite{BagGiT01} and then
studied in detail in \cite{ExnSV02,GavGiSV04,16,17,GitTySV09}. It is
important to stress that in contrast to the pure AB field case, where
particles effectively interact with the solenoid for a finite short time,
the particles moving in MSF interact with the solenoid permanently. This
opens more possibilities to study such an interaction in a number of
corresponding real physical situations. For example, using these solutions
the AB effect in cyclotron and synchrotron radiations was calculated in \cite%
{BagGiL01,BagGiL01a,BagGiT02}. Recently interest in such a superposition has
been renewed in connection with planar physics problems and the quantum Hall
effect \cite{Nambu00,ExnSV02,Lis07}. The example of the MSF stresses the
importance of studying quantum motion in superposition of the AB field with
some additional electromagnetic fields. It should be noted that exact
solutions of the Schr\"{o}dinger, Klein-Gordon and Dirac equations with the
AB field in combination with the Coulomb field and the magnetic monopole
field were studied in \cite{Villa95,18,22,23,CouPe93,BagGiT01}. Exact
solutions of the above mentioned equations with the AB field in combination
with some other electromagnetic fields were presented in \cite%
{AudSkV01,BagGiT01,28}.

The aim of the present work is to find the structure of the additional
electromagnetic fields, for which the Schr\"{o}dinger, Klein-Gordon, and
Dirac equations can be solved exactly (in what follows, we call such fields
exactly solvable additional fields), and to describe the corresponding exact
solutions.

\section{Aharonov-Bohm field and its combination with additional
electromagnetic fields}

In sections 1 and 2, we use Lorentz coordinates $x^{\nu }=(x^{0}=ct,x,y,z)$,
and cylindrical coordinates $r,\varphi $ in the $x,y$-plane$\,\,(x=r\cos
\varphi ,\,\,y=r\sin \varphi )$.

The Aharonov-Bohm field is a magnetic field $\mathbf{B}$ of an
infinitesimally thin solenoid with magnetic flux $\Phi ,$ 
\begin{equation}
B_{x}=B_{y}=0,\ B_{z}=\Phi \,\delta \left( x\right) \delta \left( y\right) =%
\frac{\Phi }{\pi r}\delta \left( r\right) ,\ r^{2}=x^{2}+y^{2},\ \Phi =%
\mathrm{const}.  \label{a1}
\end{equation}%
It can be described by the potentials $A_{\nu }^{(0)}$ of the form{\large 
\begin{eqnarray}
&&A_{\nu }^{(0)}=\left( A_{0}^{(0)},-\mathbf{A}^{(0)}\right) ,\ \ \mathbf{A}%
^{(0)}=\left( A_{x}^{(0)},\,\,A_{y}^{(0)},\,\,A_{z}^{(0)}\right) ,  \nonumber
\\
&&A_{0}^{(0)}=A_{z}^{(0)}=0,\ \ A_{x}^{(0)}=-\frac{\Phi }{2\pi }\frac{y}{%
r^{2}}=\frac{\Phi }{2\pi }\frac{\partial \varphi }{\partial x},\quad
A_{y}^{(0)}=\frac{\Phi }{2\pi }\frac{x}{r^{2}}=\frac{\Phi }{2\pi }\frac{%
\partial \varphi }{\partial y},  \label{a2} \\
&&\mathbf{A}^{(0)}=\frac{\Phi }{2\pi r}\,\mathbf{e}_{\varphi },\ \ \mathbf{e}%
_{\varphi }=-\mathbf{i}\sin \varphi +\mathbf{j}\cos \varphi \,.  \nonumber
\end{eqnarray}%
} We denote by $\mathbf{i},\mathbf{j},\mathbf{k}$ three unit orthogonal
vectors along the Cartesian axis $x,y,z,$ whereas $\mathbf{e}_{\varphi }$ and%
$\,\,\mathbf{e}_{r}$ are unit orthogonal vectors of the cylindrical
coordinate $\varphi ,\,r$ system.

An important characteristic of the AB field is the mantissa $\mu $ of the
magnetic flux $\Phi $, which is defined as follows 
\begin{equation}
\Phi =-\left( e/|e|\right) (l_{0}+\mu )\Phi _{0},\ \ 0\leqslant \mu <1,
\label{a13}
\end{equation}%
where $l_{0}$ is an integer, $\Phi _{0}=2\pi c\hbar /|e|$ is the Dirac
quantum of the magnetic flux, and $e$ the algebraic particle charge. Then 
\begin{equation}
\frac{e}{c}A_{\nu }^{(0)}=\left( 0,-\frac{e}{c}\mathbf{A}^{(0)}\right)
=\hbar (l_{0}+\mu )\partial _{\nu }\varphi ,\ \ \frac{e}{c}\mathbf{A}%
^{(0)}=-\hbar \frac{l_{0}+\mu }{r}\mathbf{e}_{\varphi }.  \label{a14}
\end{equation}

Consider a linear combination of electromagnetic potentials of the AB field
with an additional electromagnetic field. Potentials $\mathcal{A}_{\,\nu }$
of such a composite field, we represent as 
\begin{equation}
\mathcal{A}_{\,\nu }=A_{\nu }+A_{\nu }^{(0)}=\left( A_{0},-\mathbf{A}-%
\mathbf{A}^{(0)}\right) .  \label{a9}
\end{equation}%
The Schr\"{o}dinger, Klein-Gordon, and Dirac equations with such a composed
field have the form {\large 
\begin{eqnarray}
&&\mathcal{S}\Psi _{S}=0,\ \ \mathcal{S}=cP_{0}-\frac{\mathbf{P}^{2}}{2m_{0}}%
;  \label{a11a} \\
&&\mathcal{K}\Psi _{K}=0,\ \ \mathcal{K}=P^{\nu }P_{\nu
}-m_{0}^{2}c^{2}=P_{0}^{2}-\mathbf{P}^{2}-m_{0}^{2}c^{2};  \label{a11b} \\
&&\mathcal{D}\Psi _{D}=0,\ \ \mathcal{D}=\gamma ^{\nu }P_{\nu
}-m_{0}c=\gamma ^{0}P_{0}-(\boldsymbol{\gamma }\mathbf{P})-m_{0}c.
\label{a11c}
\end{eqnarray}%
} Here $m_{0}$ is the particle rest mass, $\gamma ^{\nu }$ are Dirac gamma
matrices (we use here standard representation for them), and covariant
components of the kinetic momentum operator $P_{\nu }$ are{\large 
\begin{eqnarray}
P_{\nu } &=&(P_{0},-\mathbf{P})=p_{\nu }-\frac{e}{c}\mathcal{A}_{\,\nu },\ \
P_{0}=p_{0}-\frac{e}{c}\mathcal{A}_{0},\ \ \mathbf{P}=\mathbf{p}-\frac{e}{c}%
\mbox{\boldmath${\cal
{A}}$};  \nonumber \\
p_{\nu } &=&(p_{0},-\mathbf{p})=i\hbar \partial _{\nu },\ \ p_{0}=i\hbar
\partial _{0},\ \ \mathbf{p}=-i\hbar \nabla ,  \label{a10}
\end{eqnarray}%
} where $p_{\nu }$ are covariant components of the generalized momentum
operator.

In what follows, we consider only the nontrivial AB field with nonzero
mantissa $\mu .$ As follows from (\ref{a2}), in such a case cylindrical and
spherical coordinates are physically preferable. Namely, in these
coordinates equations (\ref{a2}) with the AB field and exactly solvable
additional fields allow separation of the variables and exact solutions can
be obtained.

\section{Structure of additional electromagnetic fields. Cylindrical
coordinates}

Let us write potentials $A_{\nu }$ of the additional field in the form%
\begin{equation}
\frac{e}{c\hbar }A_{0}=f_{0}(r,\,z,\,x_{0}),\ \ \frac{e}{c\hbar }\mathbf{A}=%
\frac{f_{2}(r)}{r}\mathbf{e}_{\varphi }-f_{1}(r,\,z,\,x_{0})\mathbf{k},
\label{b1}
\end{equation}%
where $f_{k}\,\,(k=0,1,2),$ some arbitrary functions of the indicated
arguments. Thus, potentials of the AB field (\ref{a2}) can be considered as
a particular case of (\ref{b1}) for $f_{0}(r\,z,x_{0})=f_{1}(r,z,x_{0})=0,$
and$\ \ f_{2}(r)=\Phi /2\pi =\mathrm{const}.$

Electric $\mathbf{E}$ and magnetic $\mathbf{H}$ fields that correspond to
potentials (\ref{b1}) are {\large 
\begin{eqnarray}
\frac{e}{c\hbar }\mathbf{E} &=&-\partial _{r}f_{0}(r,\,z,\,x_{0})\mathbf{e}%
_{r}+[\partial _{0}f_{1}(r,\,z,\,x_{0})-\partial _{z}f_{0}(r,\,z,\,x_{0})]%
\mathbf{k},  \nonumber \\
\frac{e}{c\hbar }\mathbf{H} &=&\partial _{r}f_{1}(r,\,z,\,x_{0})\mathbf{e}%
_{\varphi }+\frac{f_{2}^{\prime }(r)}{r}\mathbf{k};\ \ \mathbf{e}_{r}=%
\mathbf{i}\cos \varphi +\mathbf{j}\sin \varphi .  \label{b2}
\end{eqnarray}%
}

In such fields, the Schr\"{o}dinger and Klein-Gordon equations have an
integral of motion%
\begin{equation}
L_{z}=[\mathbf{r}\times \mathbf{p}]_{z}=-i\hbar \,\partial _{\varphi }
\label{b2a}
\end{equation}%
and the Dirac equation has an integral of motion of the form%
\begin{equation}
J_{z}=L_{z}+\frac{\hbar }{2}\Sigma _{3}=-i\hbar \,\partial _{\varphi }+\frac{%
\hbar }{2}\Sigma _{3}.  \label{b2b}
\end{equation}

Let us look for solutions of the Schr\"{o}dinger, Klein-Gordon, and Dirac
equations with the potentials (\ref{b1}), solutions that are eigenfunctions
of the operators (\ref{b2a}) and (\ref{b2b}) the eigenvalues%
\[
L_{z}\rightarrow \hbar (l-l_{0});\,\,J_{z}\rightarrow \hbar (l-l_{0}-\frac{1%
}{2}),\ l\in \mathbb{Z}, 
\]%
respectively.

For the Schr\"{o}dinger ($S$) and Klein-Gordon ($K$) equations such
solutions have the form%
\begin{equation}
\Psi _{S,K}(x)=\exp (iQ)\psi _{S,K}(r,z,x_{0}),\ \ Q=(l-l_{0})\varphi ,
\label{b3}
\end{equation}%
where the functions $\psi _{S,K}(r,z,x_{0})$ obey the equations%
\begin{eqnarray}
&&\left\{ 2m\pi _{0}+\partial _{r}^{2}+\frac{1}{r}\partial _{r}-\frac{%
[f_{2}(r)-l-\mu ]^{2}}{r^{2}}-\pi _{3}^{2}\right\} \psi _{S}(r,z,x_{0})=0,
\label{b4} \\
&&\left\{ \pi _{0}^{2}+\partial _{r}^{2}+\frac{1}{r}\partial _{r}-\frac{%
[f_{2}(r)-l-\mu ]^{2}}{r^{2}}-\pi _{3}^{2}-m^{2}\right\} \psi
_{K}(r,z,x_{0})=0.  \label{b5a}
\end{eqnarray}%
Here%
\begin{equation}
\pi _{0}=i\partial _{0}-f_{0}(r,z,x_{0}),\ \ \pi _{3}=i\partial
_{z}-f_{1}(r,z,x_{0}),\ \ m=\frac{m_{0}c}{\hbar }.  \label{b6}
\end{equation}

For the Dirac equation ($D$), such solutions have the form 
\begin{equation}
\Psi _{D;\,l}(x^{\nu })=e^{iQ}\left( 
\begin{array}{l}
e^{-i\varphi }\psi _{1}(r,\,z,\,x_{0}) \\ 
i\psi _{2}(r,\,z,\,x_{0}) \\ 
e^{-i\varphi }\psi _{3}(r,\,z,\,x_{0}) \\ 
i\psi _{4}(r,\,z,\,x_{0})%
\end{array}%
\right) ,  \label{b7}
\end{equation}%
where the functions $\psi _{s}(r,\,z,\,x_{0}),\ s=1,\,2,\,3,\,4,$ obey the
following set of equations 
\begin{equation}
\tilde{D}\tilde{\Phi}=0,\ \tilde{\Phi}=\left( 
\begin{array}{c}
\psi _{1}(r,\,z,\,x_{0}) \\ 
\psi _{2}(r,\,z,\,x_{0}) \\ 
\psi _{3}(r,\,z,\,x_{0}) \\ 
\psi _{4}(r,\,z,\,x_{0})%
\end{array}%
\right) ,  \label{b8}
\end{equation}%
with the matrix operator $\tilde{D}$ having the form%
\begin{equation}
\overline{D}=\rho _{3}\pi _{0}+i\rho _{2}\Sigma _{3}\pi _{3}+\rho _{2}\Sigma
_{2}\left( \partial _{r}+\frac{1}{2r}\right) +i\rho _{2}\Sigma _{1}\frac{%
f_{2}(r)-l-\mu +1/2}{r}-mI.  \label{b9}
\end{equation}%
Here $\rho _{k},\ \Sigma _{k},\ k=1,\,2,\,3,$ are Dirac matrices in the
standard representation, and $\mathbb{I}$ a unit $4\times 4$ matrix.

Exact solutions of eqs. (\ref{b4}), (\ref{b5a}), and (\ref{b8}) are known
only for two types of function $f_{2}(r):$ 
\begin{equation}
a)\,\,f_{2}(r)=\gamma r;\ \ b)\,\,f_{2}(r)=\gamma r^{2},\ \ \gamma =\mathrm{%
const}.  \label{b10}
\end{equation}

\section{Klein-Gordon and Dirac equations}

Exact solutions of eqs. (\ref{b5a}) and (\ref{b8}) are known in the two
cases considered below.

\subsection{Case I:}

In this case the functions $f_{0}$ and $f_{1}$ depend on $r$ only ($%
f_{0,1}=f_{0,1}(r)$) and are linearly dependent.

Let us consider solutions of eqs. (\ref{b5a}) and (\ref{b8}) that are
eigenfunctions for both operators $i\partial _{0}$ and $i\partial _{z}$ with
the eigenvalues $k_{0}\,$and $k_{3}$ respectively. Such solutions have the
form (\ref{b3}) and (\ref{b7}) with%
\begin{equation}
Q=(l-l_{0})\varphi -k_{0}x_{0}-k_{3}z,\ \ \pi _{0}=k_{0}-f_{0}(r),\ \ \pi
_{3}=k_{3}-f_{1}(r);\ \ \psi _{K}=\psi _{K}(r),\ \ \psi _{s}=\psi _{s}(r).
\label{b11}
\end{equation}%
Equations for the functions $\psi _{K}(r)$ and$\,\,\psi _{s}(r)$ hold the
form (\ref{b5a}) and (\ref{b8}) with the natural substitution $\partial
_{r}\rightarrow d/dr$.

Let us suppose that the functions $f_{0}(r)$ and $f_{1}(r)$ are linearly
dependent. In such a case, with the help of a Lorentz transformation, which
does not changes the function $f_{2}(r)$, one can reduce the problem to the
following non-equivalent subcases:%
\begin{eqnarray*}
1)\ f_{0}(r) &=&f(r)\neq 0,\ \ f_{1}(r)=0; \\
2)\ f_{0}(r) &=&0,\ \ f_{1}(r)=f(r)\neq 0; \\
3)\ f_{0}(r) &=&\epsilon f_{1}(r)=f(r),\ \ \epsilon =\pm 1.
\end{eqnarray*}%
They are considered separately below.

\subsubsection{Subcase 1}

This subcase is characterized by the conditions%
\begin{eqnarray*}
&&f_{0}(r)=f(r)\neq 0,\ \ f_{1}(r)=0; \\
&&\frac{e}{c\hbar }\mathbf{E}=-f^{\prime }(r)\mathbf{e}_{r},\ \ \frac{e}{%
c\hbar }\mathbf{H}=r^{-1}f_{2}^{\prime }(r)\mathbf{k\ }.
\end{eqnarray*}

For such fields, the Dirac equation admits a spin integral of motion \cite%
{BagGi90} $T_{1}=m\rho _{3}\Sigma _{3}-k_{3}\rho _{1}$. We consider
solutions that are its eigenfunctions, 
\begin{equation}
T_{1}\Psi _{D;\,l}(x^{\nu })=\zeta \lambda _{1}\Psi _{D;\,l}(x^{\nu }),\ \
\zeta =\pm 1,\ \ \lambda _{1}=\sqrt{m^{2}+k_{3}^{2}}.  \label{b12}
\end{equation}%
It follows from (\ref{b12}) with account taken of (\ref{b7}) that 
\begin{eqnarray*}
&&\psi _{1}(r)=a\varphi _{1}(r),\ \ \psi _{2}(r)=-b\varphi _{2}(r),\ \ \psi
_{3}(r)=b\varphi _{1}(r),\ \ \psi _{4}(r)=-a\varphi _{2}(r); \\
&&a=\lambda _{1}+m+k_{3}+\zeta (\lambda _{1}+m-k_{3}),\ \ b=\lambda
_{1}+m-k_{3}-\zeta (\lambda _{1}+m+k_{3}).
\end{eqnarray*}%
Equations for functions $\varphi _{1}(r)$ and$\,\,\varphi _{2}(r)$ follow
from (\ref{b8}),%
\begin{eqnarray}
&&\ [k_{0}-f(r)-\zeta \lambda _{1}]\varphi _{1}(r)-\left[ \frac{%
f_{2}(r)-l-\mu }{r}-\frac{d}{dr}\right] \varphi _{2}(r)=0,  \nonumber \\
&&\left[ \frac{f_{2}(r)-l-\mu +1}{r}+\frac{d}{dr}\right] \varphi
_{1}(r)-[k_{0}-f(r)+\zeta \lambda _{1}]\varphi _{2}(r)=0.  \label{b14}
\end{eqnarray}%
Then the Klein-Gordon equation (\ref{b5a}) takes the form 
\begin{equation}
\left\{ \frac{d^{2}}{dr^{2}}+\frac{1}{r}\frac{d}{dr}+[k_{0}-f(r)]^{2}-\frac{%
[f_{2}(r)-l-\mu ]^{2}}{r^{2}}-k_{3}^{2}-m^{2}\right\} \psi _{K}(r)=0.
\label{b15}
\end{equation}

Solutions of the set (\ref{b14}) and eq. (\ref{b15}) (for nonzero $f(r)$)
are known only for%
\[
f_{2}(r)=\gamma r,\ \,f(r)=\frac{\alpha }{r};\ \ \alpha \neq 0,\,\ \gamma =%
\mathrm{const}. 
\]%
These solutions have the form 
\begin{eqnarray}
&&\varphi _{1}(r)=\sqrt{q_{1(-)}\omega _{1}(\zeta )}\,v_{1}(x)+\sqrt{%
q_{1(+)}\omega _{1}(-\zeta )}\,v_{2}(x),  \nonumber \\
&&\varphi _{2}(r)=\sqrt{q_{1(+)}\omega _{1}(\zeta )}\,v_{1}(x)+\sqrt{%
q_{1(-)}\omega _{1}(-\zeta )}\,v_{2}(x),  \nonumber \\
&&\psi _{K}(r)=v_{0}(x),  \label{b17}
\end{eqnarray}%
where 
\begin{eqnarray*}
&&q_{1(\pm )}=l+\mu -\frac{1}{2}\pm \sqrt{\left( l+\mu -\frac{1}{2}\right)
^{2}-\alpha ^{2}},\ \ E=\sqrt{m^{2}+k_{3}^{2}+\gamma ^{2}-k_{0}^{2}}, \\
&&\omega _{1}(\zeta )=\gamma \alpha -k_{0}\left( l+\mu -\frac{1}{2}\right)
+\zeta \lambda _{1}\sqrt{\left( l+\mu -\frac{1}{2}\right) ^{2}-\alpha ^{2}}%
,\ \ x=2rE.
\end{eqnarray*}%
The functions $v_{s}(x),\,(s=0,\,1,\,2)$ have similar structure, 
\begin{equation}
v_{s}(x)=AI_{p_{s},\,n_{s}}(x)+BI_{n_{s},\,p_{s}}(x),  \label{b18}
\end{equation}%
where $A$ and$\,B$ are arbitrary constants, and $I_{p,\,n}(x)$ Laguerre
functions, related to the confluent hypergeometric function $\Phi (\alpha
,\gamma ;x)$ by the relation (see \cite{GraRy94}, pp. 1072 -1073) as 
\begin{equation}
I_{p,\,n}(x)=\sqrt{\frac{{\Gamma (1+p)}}{{\Gamma (1+n)}}}{\frac{\exp (-x/2)}{%
{\Gamma (1+p-n)}}}x^{\frac{{p-n}}{2}}\Phi (-n,p-n+1;x);  \label{b19}
\end{equation}%
and the subscripts of the Laguerre functions $p_{s},\,n_{s}\,\,(s=0,\,1,\,2)$
have the form%
\begin{eqnarray*}
n_{s} &=&\frac{1}{E}\left\{ \gamma \left[ l+\mu -\frac{s(3-s)}{4}\right]
-k_{0}\alpha \right\} +\frac{(3s+1)(s-2)}{4}-\sqrt{\left[ l+\mu -\frac{s(3-s)%
}{4}\right] ^{2}-\alpha ^{2}}, \\
p_{s} &=&\frac{1}{E}\left\{ \gamma \left[ l+\mu -\frac{s(3-s)}{4}\right]
-k_{0}\alpha \right\} +\frac{(2-3s)(s-1)}{4}+\sqrt{\left[ l+\mu -\frac{s(3-s)%
}{4}\right] ^{2}-\alpha ^{2}}.
\end{eqnarray*}%
For non-negative integer $n_{s},$ the functions (\ref{b18}) are
square-integrable for $A\neq 0$ and$\,\,B=0$, and the energy $k_{0}$ is
quantized. In the case of the Dirac equation, we find%
\begin{eqnarray*}
&&k_{0}=\frac{1}{N^{2}+\alpha ^{2}}\left[ \alpha \,\gamma \left( l+\mu -%
\frac{1}{2}\right) +N\sqrt{(N^{2}+\alpha ^{2})(m^{2}+k_{3}^{2}+\gamma
^{2})-\gamma ^{2}\left( l+\mu -\frac{1}{2}\right) ^{2}}\right] , \\
&&N=n+\sqrt{\left( l+\mu -\frac{1}{2}\right) ^{2}-\alpha ^{2}};\ \ \left(
l+\mu -\frac{1}{2}\right) ^{2}\geqslant \alpha ^{2};\ \ n_{1}=n-1,\ \
n_{2}=n=0,\,1,\,2,\,...\ ,
\end{eqnarray*}%
whereas in the case of the Klein-Gordon equation we have%
\begin{eqnarray*}
&&k_{0}=\frac{1}{\overline{N}^{2}+\alpha ^{2}}\left[ \alpha \,\gamma \left(
l+\mu \right) +\overline{N}\sqrt{(\overline{N}^{2}+\alpha
^{2})(m^{2}+k_{3}^{2}+\gamma ^{2})-\gamma ^{2}\left( l+\mu \right) ^{2}}%
\right] , \\
&&\overline{N}=n+\frac{1}{2}+\sqrt{\left( l+\mu \right) ^{2}-\alpha ^{2}};\
\ \left( l+\mu \right) ^{2}\geqslant \alpha ^{2};\ \ n_{0}=n.
\end{eqnarray*}

\subsubsection{Subcase 2}

This subcase is characterized by the conditions%
\[
f_{0}(r)=0,\ \ f_{1}(r)=f(r)\neq 0;\ \ \mathbf{E}=0,\ \ \frac{e}{c\hbar }%
\mathbf{H}=f^{\prime }(r)\mathbf{e}_{\varphi }+r^{-1}f_{2}^{\prime }(r)%
\mathbf{k}\ . 
\]

Thus, we are dealing with a pure magnetic field. In such a field, the Dirac
equation admits a spin integral of motion $T_{2}=(\mathbf{\Sigma P}),$ see 
\cite{BagGi90}. Then we can impose an additional condition on the wave
function, 
\begin{equation}
(\mathbf{\Sigma P})\Psi _{D;\,l}(x^{\nu })=\zeta \lambda _{2}\Psi
_{D;\,l}(x^{\nu }),\ \ \zeta =\pm 1,\ \ \lambda _{2}=\sqrt{k_{0}^{2}-m^{2}}.
\label{b23}
\end{equation}%
Eqs. (\ref{b23}) and (\ref{b8}) are consistent if we set 
\begin{eqnarray*}
\psi _{1}(r) &=&\sqrt{k_{0}+m}\,\overline{\varphi }_{1}(r),\ \ \psi _{2}(r)=%
\sqrt{k_{0}+m}\,\overline{\varphi }_{2}(r), \\
\psi _{3}(r) &=&\zeta \sqrt{k_{0}-m}\,\overline{\varphi }_{1}(r),\ \ \psi
_{4}(r)=\zeta \sqrt{k_{0}-m}\,\overline{\varphi }_{2}(r),
\end{eqnarray*}%
in (\ref{b8}), where the functions $\overline{\varphi }_{1}(r)\,$\ and $%
\overline{\varphi }_{2}(r)$ obey equations that are similar to that of (\ref%
{b14}), 
\begin{eqnarray}
&&\ [\zeta \lambda _{2}+k_{3}-f(r)]\overline{\varphi }_{1}(r)+\left[ \frac{%
f_{2}(r)-l-\mu }{r}-\frac{d}{dr}\right] \overline{\varphi }_{2}(r)=0, 
\nonumber \\
&&\left[ \frac{f_{2}(r)-l-\mu +1}{r}+\frac{d}{dr}\right] \overline{\varphi }%
_{1}(r)+[\zeta \lambda _{2}-k_{3}+f(r)]\overline{\varphi }_{2}(r)=0.
\label{b25}
\end{eqnarray}

Now, the Klein-Gordon equation (\ref{b5a}) takes the form%
\begin{equation}
\left\{ \frac{d^{2}}{dr^{2}}+\frac{1}{r}\frac{d}{dr}-[k_{3}-f(r)]^{2}-\frac{%
[f_{2}(r)-l-\mu ]^{2}}{r^{2}}+k_{0}^{2}-m^{2}\right\} \psi _{K}(r)=0.
\label{b26}
\end{equation}

Equations (\ref{b25}) and (\ref{b26}) have exact solutions (for nonzero $%
f(r) $) only for%
\[
f(r)=\frac{\alpha }{r},\ \ f_{2}(r)=\gamma r;\ \ \alpha \neq 0,\,\gamma =%
\mathrm{const}. 
\]%
Such solutions have the form (they are similar to those in (\ref{b17}) 
\begin{eqnarray}
&&\overline{\varphi }_{1}(r)=\sqrt{q_{2(-)}\omega _{2}(\zeta )}%
\,v_{1}(x)-\zeta \sqrt{q_{2(+)}\omega _{2}(-\zeta )}\,v_{2}(x),  \nonumber \\
&&\overline{\varphi }_{2}(r)=\sqrt{q_{2(+)}\omega _{2}(\zeta )}%
\,v_{1}(x)+\zeta \sqrt{q_{2(-)}\omega _{2}(-\zeta )}\,v_{2}(x),\ \ \psi
_{K}(r)=v_{0}(x);  \nonumber \\
&&q_{2(\pm )}=\sqrt{\left( l+\mu -\frac{1}{2}\right) ^{2}+\alpha ^{2}}\pm
\left( l+\mu -\frac{1}{2}\right) ,\ \ E=\sqrt{m^{2}+k_{3}^{2}+\gamma
^{2}-k_{0}^{2}}\,,  \nonumber \\
&&\omega _{2}(\zeta )=\lambda _{2}\sqrt{\left( l+\mu -\frac{1}{2}\right)
^{2}+\alpha ^{2}}+\zeta \lbrack k_{3}\left( l+\mu -\frac{1}{2}\right)
-\gamma \alpha ],\ \ x=2rE.  \label{b27}
\end{eqnarray}%
The functions $v_{s}(x),\,(s=0,\,1,\,2)$ are given by expressions (\ref{b18}%
), with the following indices of the Laguerre functions 
\begin{eqnarray*}
n_{s} &=&\frac{1}{E}\left\{ \gamma \left[ l+\mu -\frac{s(3-s)}{4}\right]
+k_{3}\alpha \right\} +\frac{(3s+1)(s-2)}{4}-\sqrt{\left[ l+\mu -\frac{s(3-s)%
}{4}\right] ^{2}+\alpha ^{2}}\,, \\
p_{s} &=&\frac{1}{E}\left\{ \gamma \left[ l+\mu -\frac{s(3-s)}{4}\right]
+k_{3}\alpha \right\} +\frac{(2-3s)(s-1)}{4}+\sqrt{\left[ l+\mu -\frac{s(3-s)%
}{4}\right] ^{2}+\alpha ^{2}}\,.
\end{eqnarray*}

For non-negative integer $n_{s}$ the functions (\ref{b27}) are
square-integrable for $A\neq 0$ and$\,\,B=0$, and the energy $k_{0}$ is
quantized $(n_{1}=n-1,\ \ n_{2}=n_{0}=n=0,1,2,....)$ 
\begin{equation}
k_{0}^{2}=m^{2}+k_{3}^{2}+\gamma ^{2}-\left[ \frac{\gamma \left( l+\mu -%
\frac{\tau }{2}\right) +k_{3}\alpha }{n+\frac{1-\tau }{2}+\sqrt{\left( l+\mu
-\frac{\tau }{2}\right) ^{2}+\alpha ^{2}}}\right] ^{2},  \label{b29}
\end{equation}%
where $\tau =1$ for the Dirac equation and $\tau =0$ for the Klein-Gordon
equation.

\subsubsection{Subcase 3}

This subcase is characterized by the conditions%
\begin{eqnarray*}
&&f_{0}(r)=f(r)\neq 0,\ \ f_{1}(r)=\epsilon f(r),\ \ \epsilon ^{2}=1; \\
&&\frac{e}{c\hbar }\mathbf{E}=f^{\prime }(r)(\epsilon \mathbf{k}-\mathbf{e}%
_{r}),\ \ \frac{e}{c\hbar }\mathbf{H}=\epsilon f^{\prime }(r)\mathbf{e}%
_{\varphi }+r^{-1}f_{2}^{\prime }(r)\mathbf{k}\ .
\end{eqnarray*}%
In this subcase, the fields have the following properties%
\begin{equation}
\mathbf{H}=[\mathbf{n}\times \mathbf{E}]+\mathbf{n}(\mathbf{nH}),\ \ \mathbf{%
E}=-[\mathbf{n\times H}]+\mathbf{n}(\mathbf{nE}),\ \ \mathbf{n}=-\epsilon 
\mathbf{k}.  \label{b30}
\end{equation}%
If eqs. (\ref{b30}) hold, then the bispinor $\overline{\Phi }(r)$ in eq. (%
\ref{b8}) can be represented as 
\begin{eqnarray}
&&\overline{\Phi }(r)=\left( 
\begin{array}{l}
\left[ k_{0}-\epsilon k_{3}+m-\epsilon \hat{Q}\sigma _{3}\right] V(r) \\ 
\left[ \epsilon (k_{0}-\epsilon k_{3}-m)\sigma _{3}-\hat{Q}\right] V(r)%
\end{array}%
\right) ,\ \ V(r)=\left( 
\begin{array}{c}
v_{1}(r) \\ 
v_{2}(r)%
\end{array}%
\right) ,  \nonumber \\
&&\hat{Q}=\sigma _{1}\frac{f_{2}(r)-l-\mu +1/2}{r}-i\sigma _{2}\left( \frac{d%
}{dr}+\frac{1}{2r}\right) ,  \label{b31}
\end{eqnarray}%
see \cite{BagGi90}. The functions $v_{1}(r),\,v_{2}(r)$ and $\psi
_{K}(r)=v_{0}(r)$ obey the equations 
\begin{eqnarray}
&&\left\{ \frac{d^{2}}{dr^{2}}+\frac{1}{r}\frac{d}{dr}-2(k_{0}-\epsilon
k_{3})f(r)-\frac{[f_{2}(r)-l-\mu +\tau _{s}]^{2}}{r^{2}}+\delta _{s}\frac{%
f_{2}^{\prime }(r)}{r}+k_{0}^{2}-k_{3}^{2}-m^{2}\right\} v_{s}(r)=0, 
\nonumber \\
&&s=0,\,1,\,2;\ \ \tau _{s}=s(2-s);\ \ \delta _{s}=\frac{s(5-3s)}{2}.
\label{b32}
\end{eqnarray}

Equations (\ref{b32}) have exact solutions for the following two types of
the functions $f(r)$ and $f_{2}(r)$:

a)%
\begin{eqnarray}
&&f_{2}(r)=\gamma r,\ \ f(r)=\frac{\alpha }{r}+\frac{\beta }{r^{2}}; 
\nonumber \\
&&x=2rE,\ \ \alpha ,\beta ,\gamma =const;  \nonumber \\
&&E=\sqrt{m^{2}+k_{3}^{2}+\gamma ^{2}-k_{0}^{2}};  \label{b33}
\end{eqnarray}

b)%
\begin{eqnarray}
&&f_{2}(r)=\gamma r^{2},\ \ f(r)=\alpha r^{2}+\frac{\beta }{r^{2}}; 
\nonumber \\
&&x=r^{2}E_{0},\ \ \alpha ,\beta ,\gamma =\mathrm{const};  \nonumber \\
&&E_{0}=\sqrt{\gamma ^{2}+2\alpha (k_{0}-\epsilon k_{3})}.  \label{b33a}
\end{eqnarray}%
As before, the functions $v_{s}(x),\,(s=0,\,1,\,2)$ are given by expressions
(\ref{b18}), where the subindices of the Laguerre functions $n_{s},\,p_{s}$
have the form

a)%
\begin{eqnarray}
p_{s} &=&\frac{b_{s}}{E}-\frac{1}{2}+a_{s},\ \ n_{s}=\frac{b_{s}}{E}-\frac{1%
}{2}-a_{s},  \nonumber \\
b_{s} &=&\gamma (l+\mu -\tau _{s}+\frac{\delta _{s}}{2})-\alpha
(k_{0}-\epsilon k_{3}),  \nonumber \\
a_{s} &=&\sqrt{(l+\mu -\tau _{s})^{2}+2\beta (k_{0}-\epsilon k_{3})};
\label{b34}
\end{eqnarray}

b)%
\begin{eqnarray}
p_{s} &=&\frac{\tilde{b}_{s}}{4E_{0}}-\frac{1}{2}+\frac{a_{s}}{2},\ \ n_{s}=%
\frac{\tilde{b}_{s}}{4E_{0}}-\frac{1}{2}-\frac{a_{s}}{2},  \nonumber \\
\tilde{b}_{s} &=&2\gamma (l+\mu -\tau _{s}+\delta
_{s})+k_{0}^{2}-k_{3}^{2}-m^{2},  \nonumber \\
a_{s} &=&\sqrt{(l+\mu -\tau _{s})^{2}+2\beta (k_{0}-\epsilon k_{3})}.
\label{b35}
\end{eqnarray}

For non-negative integer $n_{s},$ the functions (\ref{b33}) are
square-integrable for $A\neq 0$ and$\,\,B=0$, and the energy $k_{0}$ is
quantized $(n_{1}=n-1,\ \ n_{2}=n_{0}=n=0,\,1,\,2,\,....)$. In the general
case an explicit expression for the energy $k_{0}$ has not been obtained
previously. Such expressions can be written for fields of the type a) as a
root of an algebraic equation of power six, and for fields of the type b) as
a root of an algebraic equation of power eight.

\subsection{Case II}

In this case, the functions $f_{0}$ and $f_{1}$ depend on $z$ and$\,x_{0}$
and do not depend on $r$ ($f_{0,1}=f_{0,1}(z,\,x_{0})$).

As follows from (\ref{b2}), additional fields have the form 
\begin{equation}
\frac{e}{c\hbar }\mathbf{E}=F\mathbf{k},\ \ \frac{e}{c\hbar }\mathbf{H}=G%
\mathbf{k};\ \ F=\partial _{0}f_{1}(z,\,x_{0})-\partial _{z}f_{0}(z,x_{0}),\
\ G=f_{2}^{\prime }(r)r^{-1}\ .  \label{b36}
\end{equation}%
According to a classification presented in \cite{BagGi90}, these are
longitudinal electromagnetic fields, the electric and magnetic fields are
parallel to the axis $z$, and in addition $\mathbf{H}=\mathbf{H}\left(
r\right) $ and $\mathbf{E}=\mathbf{E}\left( z,x_{0}\right) $.

As was demonstrated in \cite{BagGi90}, for such fields, solutions of the
Klein-Gordon equation (\ref{b5a}) can be found in the form 
\begin{equation}
\psi _{K}(r,z,x_{0})=w_{0}(z,\,x_{0})v_{0}(r).  \label{b37}
\end{equation}

As to solutions of the Dirac equation (\ref{b7}), we represent the bispinor $%
\overline{\Phi }(r,\,z,\,x_{0})$ in the form 
\begin{equation}
\overline{\Phi }(r,\,z,\,x_{0})=\left( 
\begin{array}{l}
\left[ m+\pi _{0}+\pi _{3}+\hat{Q}\sigma _{3}\right] w_{1}(z,\,x_{0})V(r) \\ 
\left[ (m-\pi _{0}-\pi _{3})\sigma _{3}-\hat{Q}\right] w_{1}(z,\,x_{0})V(r)%
\end{array}%
\right) ,  \label{b38}
\end{equation}%
where%
\begin{eqnarray*}
&&\pi _{0}=i\partial _{0}-f_{0}(z,\,x_{0}),\ \pi _{3}=i\partial
_{z}-f_{1}(z,\,x_{0}), \\
&&V(r)=\left( 
\begin{array}{c}
v_{1}(r) \\ 
v_{2}(r)%
\end{array}%
\right) ,\ \ \hat{Q}=\sigma _{1}\frac{f_{2}(r)-l-\mu +1/2}{r}-i\sigma
_{2}\left( \frac{d}{dr}+\frac{1}{2r}\right) .
\end{eqnarray*}%
The functions $v_{s}(r),\,s=0,\,1,\,2,$ and functions $w_{\nu
}(z,\,x_{0}),\,\nu =0,1,$ in eqs. (\ref{b37}) and (\ref{b38}) obey the
equations%
\begin{eqnarray}
&&\left\{ \frac{d^{2}}{dr^{2}}+\frac{1}{r}\frac{d}{dr}-\frac{[f_{2}(r)-l-\mu
+s(2-s)]^{2}}{r^{2}}+\frac{s(5-3s)}{2}\frac{f_{2}^{\prime }(r)}{r}+k_{\perp
}^{2}\right\} v_{s}(r)=0,  \label{b39} \\
&&\left\{ \pi _{0}^{2}-m^{2}-\pi _{3}^{2}-k_{\perp }^{2}+i\nu \lbrack
\partial _{z}f_{0}(z,\,x_{0})-\partial _{0}f_{1}(z,\,x_{0})]\right\} w_{\nu
}(z,\,x_{0})=0,\ \ k_{\perp }^{2}=\mathrm{const}.  \label{b40}
\end{eqnarray}%
Exact solutions of eq. (\ref{b39}) (for $f_{2}(r)\neq 0$) are known in the
two cases (\ref{b10}) and for $f_{2}(r)=\gamma r,\,(\gamma =\mathrm{const})$
have the form (\ref{b18}) if we set \textbf{there} $x=2r\sqrt{\gamma
^{2}-k_{\perp }^{2}}$ and 
\begin{eqnarray*}
p_{s} &=&\frac{\gamma }{\sqrt{\gamma ^{2}-k_{\perp }^{2}}}\left[ l+\mu +%
\frac{s(s-3)}{4}\right] -\frac{1}{2}+|l+\mu -s(2-s)|, \\
n_{s} &=&\frac{\gamma }{\sqrt{\gamma ^{2}-k_{\perp }^{2}}}\left[ l+\mu +%
\frac{s(s-3)}{4}\right] -\frac{1}{2}-|l+\mu -s(2-s)|.
\end{eqnarray*}

For $\gamma ^{2}>k_{\perp }^{2},\ \ \gamma (l+\mu -\tau )\geqslant 0$ ($\tau
=1/2$ for the Dirac equation and $\tau =0$ for the Klein-Gordon equation)
the quantity $k_{\perp }^{2}$ is quantized and the wave functions (\ref{b18}%
) are square-integrable for $A\neq 0,\,\,B=0,$ 
\begin{equation}
k_{\perp }^{2}=\gamma ^{2}\left[ 1-\frac{(l+\mu -\tau )^{2}}{(n+\frac{1}{2}%
+|l+\mu |)^{2}}\right] ,\ n_{0}=n_{2}=n=0,1,2,...,\ n_{1}=n-\frac{l}{|l|}.
\label{b42}
\end{equation}

The second case $f_{2}(r)=\gamma r^{2}$ corresponds to the so-called
magnetic-solenoid field (the exactly solvable additional field is a constant
uniform magnetic fields parallel to the AB solenoid) ($G=2\gamma $ in eq. (%
\ref{b36})). For such a field, exact solutions have been studied in detail
in numerous works, see e.g. \cite{Lewis83, 16, 17, BagGiL01, BagGiL01a}, and
we do not repeat these results here.

Let us consider exact solutions of the equation (\ref{b40}). A wide class of
exact solutions can be found if both functions $f_{0,\,1}(z,\,x_{0})$ depend
on one variable only: $f_{0,\,1}=f_{0,\,1}(x)$, where either $x=z$ (in such
a case, without loss of generality, we can set $f_{0}(z)=f(x)\neq 0$,$%
\,f_{1}(z)=0$); or $x=x_{0}$ (and here also without loss of generality, we
can set $f_{0}(x_{0})=0$,$\,\ f_{1}(x_{0})=f(x)\neq 0$). If $x=z$, then we
can select $w_{\nu }(x,\,x_{0})$ as an eigenfunction of $i\partial _{0}$, in
this case we find%
\begin{eqnarray}
&&i\partial _{0}w_{\nu }(x,x_{0})=k_{0}w_{\nu }(x,x_{0})\Rightarrow w_{\nu
}(x,x_{0})=\exp (-ik_{0}x_{0})w_{\nu }(x),  \nonumber \\
&&\left\{ \frac{d^{2}}{dx^{2}}+[k_{0}-f(x)]^{2}-m^{2}-k_{\perp }^{2}+i\nu
f^{\prime }(x)\right\} w_{\nu }(x)=0.  \label{b43}
\end{eqnarray}%
If $x=x_{0}$, then $w_{\nu }(z,\,x)$ can be selected as an eigenfunction of $%
i\partial _{z}$. In this case we find%
\begin{eqnarray}
&&i\partial _{z}w_{\nu }(z,x)=k_{3}w_{\nu }(z,x)\Rightarrow w_{\nu
}(z,\,x)=\exp (-ik_{3}z)w_{\nu }(x),  \nonumber \\
&&\left\{ \frac{d^{2}}{dx^{2}}+[k_{3}-f(x)]^{2}+m^{2}+k_{\perp }^{2}+i\nu
f^{\prime }(x)\right\} w_{\nu }(x)=0.  \label{b44}
\end{eqnarray}

Equations (\ref{b43}) and (\ref{b44}) are one-dimensional Schr\"{o}dinger
equations. Their exact solutions exist for the following functions $f(x):$%
\begin{eqnarray}
\,f(x) &=&\alpha \,x;\ \,f(x)=\alpha /x;\ \,f(x)=\alpha \exp (\beta x);\  
\nonumber \\
\,f(x) &=&\alpha \,\tan (\beta x);\ \,f(x)=\alpha \tanh (\beta x);\
\,f(x)=\alpha \coth (\beta x),  \label{b45}
\end{eqnarray}%
where $\alpha ,\beta =\mathrm{const.\ }$Such solutions are well-known, see
e.g. \cite{BagGi90}.

In two cases when the function $f_{0,\,1}(z,x_{0})$\ depends essentially on
both arguments $z,\,x_{0},$ one can find exact solutions of eq. (\ref{b40}).

1) Let us set $f_{0}(z,x_{0})=f_{1}(z,x_{0})=\frac{1}{2}f(\xi )$,$\ \xi
=x_{0}-z$; in this case $F=f^{\prime }(\xi )$ in eq. (\ref{b36}). Then $%
w_{\nu }(z,x_{0})$ can be selected as an eigenfunction of $i(\partial
_{0}+\partial _{z})$ with the eigenvalue $\lambda $, which implies%
\begin{eqnarray}
&&w_{\nu }(z,x_{0})=[\lambda -f(\xi )]^{-\frac{1+\nu }{2}}\exp (iS), 
\nonumber \\
&&S=-\frac{1}{2}\left[ \lambda (x_{0}+z)+(m^{2}+k_{\perp }^{2})\int \frac{%
d\xi }{\lambda -f(\xi )}\right] .  \label{b46}
\end{eqnarray}

2) Let us set $f_{0}(z,\,x_{0})=-\frac{z}{\overline{\xi }}f(\overline{\xi }%
),\ \ f_{1}(z,x_{0})=\frac{x_{0}}{\overline{\xi }}f(\overline{\xi });\ \ 
\overline{\xi }=x_{0}^{2}-z^{2}$; in this case $F=2f^{\prime }(\overline{\xi 
})$ in eq. (\ref{b36}). Then $w_{\nu }(z,x_{0})$ can be selected as an
eigenfunction of the operator $\hat{q}=i(z\partial _{0}+x_{0}\partial _{z})$
(this operator is an integral of motion for the Klein-Gordon and Dirac
equations) with the eigenvalue $\lambda $, which implies 
\begin{equation}
\hat{q}w_{\nu }(z,x_{0})=\lambda w_{\nu }(z,x_{0})\Rightarrow w_{\nu
}(z,x_{0})=\left( \frac{x_{0}-z}{x_{0}+z}\right) ^{i\frac{\lambda }{2}%
}w_{\nu }(\overline{\xi }).  \label{b47}
\end{equation}%
Substituting (\ref{b47}) into (\ref{b40}), we find an equation for the
function $w_{\nu }(\overline{\xi }),$%
\begin{equation}
4\overline{\xi }^{2}w_{\nu }^{\prime \prime }(\overline{\xi })+4\overline{%
\xi }w_{\nu }^{\prime }(\overline{\xi })+\left\{ [\lambda -f(\overline{\xi }%
)]^{2}+\overline{\xi }[m^{2}+k_{\perp }^{2}+2i\nu f^{\prime }(\overline{\xi }%
)]\right\} w_{\nu }(\overline{\xi })=0.  \label{b48}
\end{equation}%
Equation (\ref{b48}) can be solved exactly in two cases:

a)%
\[
\ f(\overline{\xi })=\alpha \overline{\xi },\ \ \alpha =\mathrm{const}, 
\]%
which corresponds to a constant and uniform electric field;

b)%
\[
\ f(\overline{\xi })=\alpha \sqrt{|\overline{\xi }|},\ \ \alpha =\mathrm{%
const}. 
\]
In these cases solutions have the form (\ref{b18}), 
\begin{equation}
w_{\nu }(\overline{\xi })=AI_{p,\,n}(x)+BI_{n,\,p}(x),\ \ A,\,B=\mathrm{const%
},  \label{b49}
\end{equation}%
where one has to set respectively

a)%
\[
f(\overline{\xi })=\alpha \overline{\xi },\ \alpha =\mathrm{const},\ \
x=-i\alpha \overline{\xi },\ \ p=\frac{i(m^{2}+k_{\perp }^{2})-2\alpha
(1+\nu )}{4\alpha },\ n=p-i\lambda ; 
\]%
and

b)%
\begin{eqnarray*}
&&f(\overline{\xi })=\alpha \sqrt{|\overline{\xi }|},\ \ \alpha =\mathrm{%
const},\ \ x=2i\sqrt{\alpha ^{2}|\overline{\xi }|+\overline{\xi }%
(m^{2}+k_{\perp }^{2})}, \\
&&p=\frac{\alpha (\nu +2i\lambda )}{\sqrt{\alpha ^{2}+\varepsilon
(m^{2}+k_{\perp }^{2})}}-1/2+i\lambda ,\ \ n=p-2i\lambda ,\ \ \varepsilon =%
\overline{\xi }/|\overline{\xi }|\,.
\end{eqnarray*}

Thus, with the above consideration, all the exactly solvable additional
electromagnetic fields in the cylindric coordinates have been exhausted.

\section*{Schr\"{o}dinger equation}

Let us consider the Schr\"{o}dinger equation (\ref{b4}). Exact solutions of
this equation are known also only for two types of function $f_{2}(r)$.

Let us suppose that additional fields can depend on the constant $m$ (on the
particle mass). Then:

a) For \thinspace $f_{2}(r)=\gamma r$ one can find exact solutions for%
\begin{equation}
f_{0}(r)=\frac{\alpha }{r}+\frac{\delta }{r^{2}}+\frac{2\lambda }{r^{3}}%
\left( \beta -\frac{m\lambda }{r}\right) ,\ \ f_{1}(r)=\frac{\beta }{r}-%
\frac{2m\lambda }{r^{2}};\ \ \alpha ,\,\beta ,\,\delta ,\,\lambda =\mathrm{%
const.}  \label{b51}
\end{equation}%
The general solution of the Schr\"{o}dinger equation (\ref{b4}) can be
expressed via the Laguerre functions and has the form 
\begin{eqnarray}
&&\psi _{S}(r)=AI_{p,\,n}(x)+BI_{n,\,p}(x),\ \ x=2r\sqrt{E},\ \ p=\frac{b}{%
\sqrt{E}}-\frac{1}{2}+\sqrt{a},\ \ n=\frac{b}{\sqrt{E}}-\frac{1}{2}-\sqrt{a};
\nonumber \\
&&a=(l+\mu )^{2}+\beta ^{2}+2m\delta +4m\lambda k_{3},\ \ b=\gamma (l+\mu
)+\beta k_{3}-\alpha \,m,\ \ E=\gamma ^{2}+k_{3}^{2}-2mk_{0},  \label{b52}
\end{eqnarray}%
where $A$ and $B$ are arbitrary constants. For $a\geqslant 0$ and
non-negative integer $n=0,\,1,\,2,\,...,$ the functions (\ref{b9}) and (\ref%
{b10}) are square-integrable at $B=0,$ and the non-relativistic particle
energy $k_{0}$ is quantized,%
\begin{equation}
\psi _{S}(r)=A\exp (-x/2)x^{\sqrt{a}}L_{n}^{2\sqrt{a}}(x),\ \ k_{0}=\frac{1}{%
2m}\left[ \gamma ^{2}+k_{3}^{2}-\frac{4b^{2}}{(2n+1+2\sqrt{a})^{2}}\right] ,
\label{b53}
\end{equation}%
where $L_{n}^{\alpha }(x)$ are Laguerre polynomials (see \cite{GraRy94}, eq.
8.970.1).

b) \thinspace $\,f_{2}(r)=\gamma r^{2}$ one can find exact solutions for $%
\alpha ,\,\beta ,\,\gamma ,\,\delta ,\,\lambda =\mathrm{const,}$ and 
\begin{equation}
f_{0}(r)=\alpha r^{2}+\frac{\beta }{r^{2}}-2m\left( \frac{\lambda ^{2}}{r^{4}%
}+\delta ^{2}r^{4}\right) ,\ \ f_{1}(r)=-2m\left( \frac{\lambda }{r^{2}}%
+\delta r^{2}\right) .  \label{b54}
\end{equation}

As in the previous case, the general solution of the Schr\"{o}dinger
equation (\ref{b4}) can be expressed via the Laguerre functions and has the
form%
\[
\psi _{S}(r)=AI_{p,\,n}(x)+BI_{n,\,p}(x), 
\]%
where $A$ and $B$ are arbitrary constants and%
\begin{eqnarray}
x &=&\sqrt{b}\,r^{2},\ \ p=\frac{E}{4\sqrt{b}}-\frac{1}{2}+\frac{\sqrt{a}}{2}%
,\ \ n=\frac{E}{4\sqrt{b}}-\frac{1}{2}-\frac{\sqrt{a}}{2},  \nonumber \\
a &=&(l+\mu )^{2}+2\beta m+4m\lambda k_{3},\ \ b=\gamma ^{2}+2\alpha
m+4m\delta k_{3},  \nonumber \\
E &=&2mk_{0}+2\gamma (l+\mu )-k_{3}^{2}-8\delta \lambda m^{2}.  \label{b55}
\end{eqnarray}

For $a\geqslant 0,\,b>0$ and non-negative integer $n,$ the function $\psi
_{S}(r)$ is square-integrable and the non-relativistic particle energy $%
k_{0} $ is quantized, 
\begin{eqnarray*}
&&k_{0}=\frac{1}{2m}\left[ 2\sqrt{b}(2n+1+\sqrt{a})+k_{3}^{2}-2\gamma (l+\mu
)+8\delta \lambda m^{2}\right] , \\
&&\psi _{S}(r)=A\exp (-x/2)x^{\frac{\sqrt{a}}{2}}L_{n}^{\sqrt{a}}(x).
\end{eqnarray*}

If additional fields do not depend on $m$, then one has to set $\lambda =0$
in eqs. (\ref{b51}) and (\ref{b52}), and $\delta =\lambda =0$ in eqs. (\ref%
{b54}) and (\ref{b55}).

For the fields of case II, solutions of the Schr\"{o}dinger equation (\ref%
{b4}) can be written as 
\begin{equation}
\psi _{S}(r,z,x_{0})=w_{S}(z,x_{0})v_{0}(r),  \label{b57}
\end{equation}%
where the function $v_{0}(r)$ is a solution of eq. (\ref{b39}) (for $s=0$).
Exact solutions are known only for the functions (\ref{b10}). The function $%
w_{S}(z,\,x_{0})$ is a solution of the equation 
\begin{equation}
\left\{ 2m[i\partial _{0}-f_{0}(z,x_{0})]-k_{\perp }^{2}-[i\partial
_{z}-f_{1}(z,x_{0})]^{2}\right\} w_{S}(z,x_{0})=0.  \label{b58}
\end{equation}

If $f_{0}(z,\,x_{0})=f(z)$ and$\ f_{0}(z,\,x_{0})=0$, the function $%
w_{S}(z,\,x_{0})$ can be selected as an eigenfunction for the operator $%
i\partial _{0},$%
\begin{equation}
i\partial _{0}w_{S}(z,\,x_{0})=k_{0}w_{S}(z,x_{0})\Rightarrow
w_{S}(z,x_{0})=\exp (-ik_{0}x_{0})w_{S}(z).  \label{b59}
\end{equation}%
Taking into account (\ref{b58}), we obtain for $w_{S}(z)$ the
one-dimensional Schr\"{o}dinger equation following equation 
\begin{equation}
\left\{ 2m[k_{0}-f(z)]-k_{\perp }^{2}+\frac{d^{2}}{dz^{2}}\right\} w_{S}(z)=0
\label{b60}
\end{equation}%
already discussed above.

If $f_{0}(z,\,x_{0})=0$ and$\ f_{1}(z,\,x_{0})=f(x_{0})$ (which correspond
to a uniform electric field that depends on time), the function $%
w_{S}(z,\,x_{0})$ can be selected as an eigenfunction for the operator $%
i\partial _{z},$%
\begin{equation}
i\partial _{z}w_{S}(z,\,x_{0})=k_{3}w_{S}(z,\,x_{0})\Rightarrow
w_{S}(z,\,x_{0})=\exp (-ik_{3}z)w_{S}(x_{0}).  \label{b61}
\end{equation}%
Taking into account (\ref{b58}), we obtain for $w_{S}(x_{0})$ the equation 
\begin{equation}
\left\{ 2im\partial _{0}-k_{\perp }^{2}-[k_{3}-f(x_{0})]^{2}\right\}
w_{S}(x_{0})=0.  \label{b62}
\end{equation}%
Its exact solution can be found for an arbitrary $f(x_{0}),$ 
\begin{equation}
w_{S}(x_{0})=\exp [-iS(x_{0})],\ \ S(x_{0})=\frac{k_{\perp }^{2}}{2m}%
x_{0}+\int \frac{[k_{3}-f(x_{0})]^{2}}{2m}\,dx_{0}.  \label{b63}
\end{equation}

For arbitrary functions $f_{0,\,1}(z,\,x_{0}),$ equation (\ref{b58}) can be
reduced to the one-dimensional Schr\"{o}dinger equation 
\begin{eqnarray}
&&i\partial _{0}\psi (z,\,x_{0})=\mathcal{H}\psi (z,\,x_{0}),\ \ \mathcal{H}%
=-\frac{1}{2m}\frac{d^{2}}{dz^{2}}+\mathcal{V}(z,\,x_{0}),  \label{b65} \\
&&\mathcal{V}(z,\,x_{0})=f_{0}(z,\,x_{0})-\int \left[ \partial
_{0}f_{1}(z,\,x_{0})\right] dz  \nonumber
\end{eqnarray}%
with the help of the substitution 
\begin{equation}
w_{S}(z,\,x_{0})=\exp [-i\varphi (z,x_{0})]\psi (z,\,x_{0}),\ \ \varphi
(z,x_{0})=\frac{k_{\perp }^{2}}{2m}x_{0}+\int f_{1}(z,\,x_{0})dz.
\label{b64}
\end{equation}%
We note that the potential in such an equation depends on time.

We believe that there exist new kinds of functions $f_{0,\,1}(z,\,x_{0})$
for which exact solutions can be found. The method of constructing such
functions and corresponding solutions is developed in \cite{30,31}.

Thus, with the above consideration, all the exactly solvable additional
electromagnetic fields in the case of the Schr\"{o}dinger equation in the
cylindric coordinates have been exhausted.

\section{Structure of additional electromagnetic fields. Spherical
coordinates}

In the spherical coordinates $r,\,\theta ,\,\varphi ,$ 
\begin{equation}
x=r\sin \theta \cos \varphi ,\ \ y=r\sin \theta \sin \varphi ,\ \ z=r\cos
\theta  \label{c1}
\end{equation}%
potentials of the AB field (\ref{a2}) have the form%
\begin{equation}
A_{\nu }^{(0)}=\left( A_{\,0}^{(0)},-\mathbf{A}^{(0)}\right) ,\ \mathbf{A}%
^{(0)}=\frac{\Phi }{2\pi r\sin \theta }\mathbf{e}_{\varphi },\ \mathbf{e}%
_{\varphi }=-\mathbf{i}\sin \varphi +\mathbf{j}\cos \varphi \,.  \label{c2}
\end{equation}

Potentials $A_{\nu }$ of additional fields we write as 
\begin{equation}
\frac{e}{c\hbar }A_{0}=f_{0}(r),\ \frac{e}{c\hbar }\mathbf{A}=\frac{%
f_{1}(\cos \theta )}{r\sin \theta }\mathbf{e}_{\varphi },  \label{c3}
\end{equation}%
where $f_{k}\,\,(k=0,1)$ are arbitrary functions of their arguments. Thus,
the potentials of the AB field (\ref{a2}) can be considered as a particular
case of the additional field (\ref{c3}) for $f_{0}(r)=0$ and$\ f_{1}(\cos
\theta )=\Phi /2\pi =\mathrm{const}.$

Electric and magnetic field that corresponds to (\ref{c3}) have the form%
\begin{equation}
\frac{e}{c\hbar }\mathbf{E}=-f_{0}^{\prime }(r)\mathbf{e}_{r},\ \frac{e}{%
c\hbar }\mathbf{H}=-\frac{f_{1}^{\prime }(\cos \theta )}{r^{2}}\mathbf{e}%
_{r},\ \ \mathbf{e}_{r}=\sin \theta (\mathbf{i}\cos \varphi +\mathbf{j}\sin
\varphi )+\mathbf{k}\cos \theta \,.  \label{c4}
\end{equation}

A complete procedure of variable separation and finding complete sets of
integrals of motion for the Schr\"{o}dinger, Klein-Gordon and Dirac
equations with potentials (\ref{c3}) is presented in detail in \cite{BagGi90}%
.

One can find exact solutions of the Klein-Gordon and Dirac equations for 
\begin{equation}
f_{1}(\cos \theta )=\alpha \cos \theta +\beta ,\ \ f_{0}(r)=\frac{\gamma }{r}%
,\ \ \alpha ,\beta ,\gamma =\mathrm{const}.  \label{c5}
\end{equation}%
In this case the complete external electromagnetic field is a combination of
the AB field with the Coulomb field and the field of a magnetic monopole.
The corresponding exact solutions were studied in the works \cite%
{Villa95,18,22,23,CouPe93,BagGiT01}.

As to the Schr\"{o}dinger equation, its solutions can be found for a more
general potentials%
\begin{equation}
f_{1}(\cos \theta )=\alpha \cos \theta +\beta ,\ \ f_{0}(r)=\frac{\gamma }{r}%
+\frac{\delta }{r^{2}},\ \ \alpha ,\beta ,\gamma ,\delta =\mathrm{const}.
\label{c6}
\end{equation}%
The Schr\"{o}dinger equation with potentials (\ref{c6}) can be trivially
identified with the Klein-Gordon equation with the potentials (\ref{c5})
and, therefore, does not need a separate consideration.

\section{Conclusion}

Exact solutions of quantum mechanical wave equations for a combination of
Aharonov-Bohm field with additional external electromagnetic fields has an
undoubted physical interest. The additional fields can emphasize or even
reinforce some specific manifestations of the Aharonov--Bohm effect. As was
already said, exact solutions in the case when the additional field is a
constant and uniform magnetic field, revealed non-trivial manifestations of
the Aharonov-Bohm effect in the synchrotron radiation.

Only some of additional fields and the corresponding exact solutions, which
are considered above, have been sufficiently studied until now. Namely, a
constant and uniform magnetic field that is parallel to the Aharonov-Bohm
solenoid, a static spherically symmetrical electric field (in particular
Coulomb field), and a field of a magnetic monopole.

In this work we have demonstrated that aside from these known cases, there
are broad classes of additional external fields, for which exact solutions
exist for both relativistic and non relativistic wave equations. Among these
new additional fields we have physically interesting electric fields acting
during a finite time, or localized in a finite region of space (see the
potentials in (\ref{b45}). The corresponding exact solutions can be used in
non perturbative calculations of different processes in QED with unstable
vacuum (see. \cite{FGS}) and, therefore, to study the Aharonov-Bohm effect
in such processes. There are additional time-dependent uniform and isotropic
electric fields that allow exact solutions of the Schr\"{o}dinger equation.
It should be also noted that in the relativistic case these are additional
electric fields propagating along the Aharonov-Bohm solenoid with arbitrary
electric pulse shape (\ref{b46}).

Obtained results give a possibility to investigate precisely manifestations
of the Aharonov-Bohm effect in the background of a wide class of additional
electromagnetic fields.

\section*{Acknowledgments}

The work of VGB was partially supported by the Federal Targeted Program
"Scientific and scientific - pedagogical personnel of innovative Russia",
contract\thinspace\ 02.740.11.0238; N. P789; DMG acknowledges the permanent
support of FAPESP and CNPq.


\begin{thebibliography}{99}
\bibitem{AhaBo59} Y. Aharonov and D. Bohm, \emph{Significance of
electromagnetic potentials in quantum theory,} Phys. Rev. \textbf{115}
(1959) 485-491

\bibitem{EhrSi49} W.~Ehrenberg and R.E.~Siday, \emph{The Refractive Index in
Electron Optics and the Principles of Dynamics}, Proc. Phys. Soc. Lond., B 
\textbf{62} (1949) 8-212

\bibitem{WuYa75} T.T. Wu and C.N. Yang, \emph{Concept of nonintegrable phase
factors and global formulation of gauge fields}, Phys. Rev. D \textbf{12}
(1975) 3845-3857

\bibitem{PerVi89} U. Percoco, V.M. Villalba, \emph{Aharonov - Bohm Effect
for a Relativistic Dirac Electron}, Phys. Lett. A \textbf{140}(3) (1989)
105-107

\bibitem{Hagen91} C.R. Hagen, \emph{Aharonov-Bohm Scattering of Particles
with Spin}, Phys. Rev. Letters, \textbf{64 (}1990) 503-506; \emph{Spin
dependence of the Aharonov--Bohm effect}, Int. J. Mod. Phys. A \textbf{6}
(1991) 3119-3149

\bibitem{Hagen93} C.R. Hagen, \emph{Effects of nongauge potentials on the
spin-1/2 Aharonov-Bohm problem}, Phys. Rev. \textbf{D 48} (1993) 5935-5939

\bibitem{VGS91} S.A. Voropaev, D.V. Galtsov, and D.A. Spasov, \emph{Bound
states for fermions in the gauge Aharonov-Bohm field}, Phys. Lett. B\textbf{%
\ 267} (1991) 91-94

\bibitem{Nambu00} Y. Nambu, \emph{The Aharonov-Bohm problem revisited},
Nucl. Phys. B \textbf{579} (2000) 590-616; M. Hirokawa and O. Ogurisu, \emph{%
\ Ground state of a spin-1/2 charged particle in a two-dimensional magnetic
field}, J. Math. Phys., \textbf{42} (2001) 3334-3343

\bibitem{CNP92} F.A.B. Coutinho, Y. Nogami, and J.F. Perez, \emph{%
Self-adjoint extensions of the Hamiltonian for a charged-particle in the
presence of a thread of magnetic-flux}, Phys. Rev. A \textbf{46} (1992)
6052-6055; \emph{Self-adjoint extensions of the Hamiltonian for a charged
spin-1/2 particle in the Aharonov-Bohm field}, Journ. Phys. A \textbf{27}
(1994) 6539-6550

\bibitem{CouPe94} F.A.B. Coutinho and J.F. Perez, \emph{Helicity
conservation in the Aharonov-Bohm scattering of Dirac Particles}, Phys. Rev.
D \textbf{49} (1994) 2092-2097

\bibitem{FalPi01} H. Falomir and P.A.G. Pisani, \emph{Hamiltonian
self-adjoint extensions for (2+1)-dimensional Dirac particles,} Journ. Phys.
A \textbf{34} (2001) 4143-4154

\bibitem{Lewis83} R.R.~Lewis, \emph{Aharonov-Bohm effect for trapped ions,}
Phys. Rev. A \textbf{28} (1983) 1228-1236

\bibitem{BagGiT01} V.G. Bagrov, D.M. Gitman, and V.B. Tlyachev, \emph{%
Solutions of relativistic wave equations in superpositions of Aharonov-Bohm,
magnetic, and electric fields}, J. Math. Phys. \textbf{42 }(2001) 1933-1959

\bibitem{ExnSV02} P. Exner, P. \v{S}t'ovi\v{c}ek, and P. Vyt\v{r}as, \emph{%
Generalized boundary conditions for the Aharonov-Bohm effect combined with a
homogeneous magnetic field}, J. Math. Phys., \textbf{43} (2002) 2151-2168

\bibitem{GavGiSV04} S.P. Gavrilov, D.M. Gitman, A.A. Smirnov, and B.L.
Voronov, \emph{Dirac fermions in a magnetic-solenoid field, }hep-th/0308093;
\textquotedblright Focus on Mathematical Physics Research\textquotedblright\
Ed. by Charles V. Benton (Nova Science Publishers, New York, 2004) 131-168

\bibitem{16} V.G. Bagrov, S.P. Gavrilov, D.M. Gitman, and D.P. Meira Filho, 
\emph{Coherent states of non-relativistic electron in magnetic--solenoid
field, }Journ. Physics. A\textbf{43} (2010) 3540169 (10 pages)

\bibitem{17} V.G. Bagrov, S.P. Gavrilov, D.M. Gitman, and D.P. Meira Filho, 
\emph{Coherent and semiclassical states in magnetic field in the presence of
the Aharonov-Bohm solenoid}, J. Phys. A: Math. Theor. \textbf{44} (2011)
055301 (35 pages).

\bibitem{GitTySV09} D.M. Gitman, I.V. Tyutin, A. Smirnov, and B.L. Voronov, 
\emph{Self-adjoint Schr\"{o}dinger and Dirac operators with Aharonov-Bohm
and magnetic-solenoid fields}, arXiv:0911.0946 [quant-ph]

\bibitem{BagGiL01} V.G. Bagrov, D.M. Gitman, A. Levin, and V.B. Tlyachev, 
\emph{Impact of Aharonov-Bohm Solenoid on Particle Radiation in Magnetic
Field}, Mod. Phys. Lett. A \textbf{16} (2001) 1171-1179

\bibitem{BagGiL01a} V.G. Bagrov, D.M. Gitman, A. Levin, and V.B. Tlyachev, 
\emph{Aharonov-Bohm effect in cyclotron and synchrotron radiations}, Nucl.
Phys. B \textbf{605} (2001) 425-454

\bibitem{BagGiT02} V.G. Bagrov, D.M. Gitman, and V.B. Tlyachev, \emph{%
l-dependence of particle radiation in magnetic-solenoid field and
Aharonov-Bohm effect}, Int. Journ. Mod. Phys. A\textbf{17 }(2002) 1045-1048

\bibitem{Lis07} O. Lisovyy, \emph{Aharonov-Bohm effect on the Poincare disc}%
, J. Math. Phys. \textbf{48} (2007) 052112-17

\bibitem{Villa95} V.M. Villalba, \emph{Exact solutions of the Dirac equation
for a Coulomb and scalar potential in the presence of an Aharonov-Bohm and
magnetic monopole fields}, J. Math. Phys. \textbf{36} (1995) 3332-3344

\bibitem{18} Le Van Hoang, Ly Xuan Hai, L. I. Komarov, T.S. Romanova, \emph{%
Relativistic Analogy of the Aharonov -- Bohm Effect in the Presence of
Coulomb Field and Magnetic Charge,} Journal of Physics. A\textbf{25 (}25%
\textbf{)} (1992) 6461-6469

\bibitem{22} V.R. Khalilov, \emph{Relativistic Aharonov -- Bohm Effect in
the Presence of Planar Coulomb Potentials}, Phys. Rev. A \textbf{71}(1)
(2005) 012105 (6 pages)

\bibitem{23} A.D. Alhaidari, \emph{The Three--Dimensional Dirac Oscillator
in the Presence of Aharonov--Bohm and Magnetic Monopole Potentials},
Foundations of Physics Letters \textbf{18}(7) (2005) 651-664

\bibitem{CouPe93} F.A.B. Coutinho and J.F. Perez, \emph{Boundary-conditions
in the Aharonov-Bohm scattering of Dirac particles and the effect of Coulomb
interaction}, Phys. Rev. D\textbf{\ 48} (1993) 932-939

\bibitem{AudSkV01} J. Audretsch, V. Skarzinsky, and B. Voronov, \emph{%
Elastic scattering and bound states in the Aharonov-Bohm potential
superinposed by an attractive }$\rho ^{-2}$\emph{\ potential}, Journ. Phys.
A \textbf{34} (2001) 235-250

\bibitem{28} V.G. Bagrov, M.C. Baldiotti, D.M. Gitman, and I.V. Shirokov, 
\emph{New solutions of relativistic wave equations in magnetic field and
longitudinal fields}, J. Math. Phys. \textbf{43} (2002) 2284-2295

\bibitem{BagGi90} V.G. Bagrov and D.M. Gitman, \emph{Exact Solutions of
Relativistic Wave Equations}, Math. its Appl., Sov. ser., Vol \textbf{39}
(Kluwer, Dordrecht 1990)

\bibitem{GraRy94} I.S. Gradshtein and I.W. Ryzhik, \emph{Table of Integrals,
Series, and Products} (Academic Press, New York 1994)

\bibitem{30} V.N. Shapovalov, N.B. Sukhomlin, \emph{Separation of Variables
in the Nonstationary Schr\"{o}dinger Equation}, Soviet Phys. Journ. \textbf{%
17}(12) (1974) 1718-1722

\bibitem{31} V.G. Bagrov, A.V. Shapovalov, I.V. Shirokov, \emph{Generation
on Exactly Solvable Potentials for the Nonstationary Schr\"{o}dinger Equation%
}, Soviet Theor. Math. Phys. \textbf{87}(3) (1991) 426-433

\bibitem{FGS} E.S.Fradkin, D.M. Gitman, and Sh.M. Shvartsman, \emph{Quantum
Electrodynamics with Unstable Vacuum} (Springer-Verlag, Berlin, 1991)
\end{thebibliography}
\end{document}